\documentclass[aps,twocolumn,prb,superscriptaddress,amsmath,showpacs,tightenlines]{revtex4}

\usepackage{epsfig}
\usepackage{color}
\usepackage{graphicx}

\begin{document}
\title{Detection of a charged two-level system by using the Kondo and the Fano-Kondo effects in quantum dots}

\author{Tetsufumi Tanamoto}
\affiliation{Corporate R \& D center, Toshiba Corporation,
Saiwai-ku, Kawasaki 212-8582, Japan}

\author{Yu-xi Liu}
\affiliation{Institute of Microelectronics, Tsinghua University, Beijing 100084, China}
\affiliation{Tsinghua National Laboratory for Information Science and Technology (TNList), Tsinghua University, Beijing 100084, China}

\author{Xuedong Hu}
\affiliation{Department of Physics, University at Buffalo, SUNY,
Buffalo, New York 14260-1500,USA}

\author{Franco Nori}
\affiliation{Advanced Science Institute, RIKEN, Wako-shi, Saitama 351-0198, Japan}
\affiliation{Physics Department, The University of Michigan, Ann Arbor, Michigan 48109, USA}

\date{\today}

\begin{abstract}
The Kondo effect and the Fano-Kondo effect are important phenomena that have been observed in  quantum dots (QDs). 
We theoretically investigate the transport properties of a coupled QD system 
in order to study the possibility of detecting a qubit state from the modulation of 
the conductance peak in the Kondo effect and the dip in the Fano-Kondo effect.
We show that the peak and dip of the conductance are both shifted depending on the qubit state.
In particular, we find that we can estimate the optimal point and tunneling coupling between the $|0\rangle$ and $|1\rangle$ states of the qubit 
by measuring the shift of the positions of the conductance peak and dip, 
as functions of the applied gate voltage on the qubit and the distance between the qubit and the detector.
\end{abstract}
 
\maketitle

\section{Introduction}
Nanodevices allow the observation of interesting quantum interference effects.
The Kondo effect and the Fano-Kondo effect are observed in coupled systems with discrete energy-levels 
and a continuum of states, {\it e.g.},
when a quantum dot (QD) is tunnel-coupled to leads.
The Kondo effect in a QD appears as a zero-bias peak in the conductance because of the spin singlet formation
between a localized spin and the reservoirs~\cite{Kondo}, while 
in the Fano-Kondo effect, an asymmetric line shape is observed in the density of states (DOS) and the conductance 
because of interferences in the hybrid electron states in the dot-electrode system~\cite{Fano}.

In a Kondo system, such as a QD connected to two electrodes, 
the conductance has a sharp peak, known as the Kondo peak~\cite{Wingreen,Hofstetter,Sasaki1,Sasaki3,Kogan,Kobayashi,Dagotto}; while
in Fano-Kondo systems, such as T-shaped QDs, 
the conductance has a sharp dip (Fano-Kondo dip) structure~\cite{Wu,Kang,Ueda,Gores,Sato,Rushforth,Sasaki2}, 
both of these as a function of the energy level of the QDs.
The peak and dip structures appear when the energy-level is close to the 
Fermi level of the reservoirs.

Here we investigate the Kondo effect and the Fano-Kondo effect 
using them as detectors of a capacitively-coupled two-level system, a charge qubit~\cite{Buluta}. 
We use Fig.~1(a) as a set-up for the Kondo effect and Fig.~1(b) to study the Fano-Kondo effect.
Each grey ellipse in Fig.~1 represents a QD. The source ``S", drain ``D", and QD $``d"$ are the 
``linear-shaped" detector in the Kondo geometry shown in Fig.~1(a). The T-shaped detector in Fig.~1(b) has a trap site $``c"$ 
and a proper QD $``d"$. 
We define the $|0\rangle$ state of the charge qubit when the excess charge is localized in the QD $``a"$, and 
the $|1\rangle$ state when the excess charge is localized in the QD $``b"$. 
Using the notation in Fig.~1, due to the Coulomb interaction $V_q$ between the charge qubit 
and the detectors,
the energy level of the QD $``d"$ is shifted for the Kondo detector, 
and that of the QD $``c"$ is shifted for the Fano-Kondo detector. 

The basic idea is the following: the Kondo peak and the Fano-Kondo dip will be affected by the charge state 
of the charge qubit because of the capacitive 
coupling between the charge qubit and the detectors.
Therefore, it is expected that, by analyzing the change of the Kondo peak and the Fano-Kondo dip in the conductance, 
the charge qubit state can be inferred. 
In our model, the capacitive coupling is the same as those of 
the conventional quantum point contact (QPC)~\cite{Elzerman,TanaHu}, 
and the single electron transistor (SET)~\cite{Astafiev,TanaHu2}.
In the standard QPC or SET system, only the position of the excess charge in the qubit ($|0\rangle$ or $|1\rangle$) is detected.  
What's new here is that, by analyzing the peak position of the conductance peak and dip, 
we can also estimate the tunneling coupling $\Omega$ between the $|0\rangle$ and $|1\rangle$ states of the charge qubit. 
We will also show that the shifts of the conductance peak and dip are largest 
when the energy gap between the two qubit eigenstates is smallest.
At this point, we will show that the Fano factor is smallest and we 
call this point the {\it optimal point}, 
where in general charge-noise-induced dephasing is minimized~\cite{Vion,Li,TanaHu}.

The standard method to detect the position of the excess charge in a charge qubit is by measuring 
the conductance of a single-electron transistor near a Coulomb blockade peak.  
This standard method is considered to be more robust than our method, 
because, in general, measurements of the Kondo and Fano-Kondo effects are more difficult than the 
measurement of a Coulomb blockade. 
In this respect, our method has a supplementary relationship to the standard method.
In the Kondo and Fano-Kondo regime, which emerge as a result of correlation between the localized spin and the Fermi sea, 
charge degrees of freedom are not perfectly frozen. 
Therefore, in the conventional Kondo or Fano-Kondo regime, 
the charge and spin degrees of freedom are not separated, in contrast to the pure one-dimensional Luttinger liquid 
that shows spin charge separation~\cite{LT}. 
This means that electrons in the QDs of the Kondo regime behave differently than those of the non-Kondo regime, 
as a result of the correlation between the charge and spin degrees of freedom.
It is considered that our setups provides the information about the system by the linking between 
the spin and charge degree of freedom of the QDs.

It is well-known that there are many two-level systems in various materials~\cite{Anderson,Philips} and 
our method should be able to detect those two-level systems or, in general, any systems that have two charged states. 
In addition, two-level systems based on QDs are also widely used in spin qubits~\cite{Taylor,Hu1}.
Thus, this method has a wide variety of potential applications for nanosystems.

\begin{figure}
\includegraphics[width=8cm]{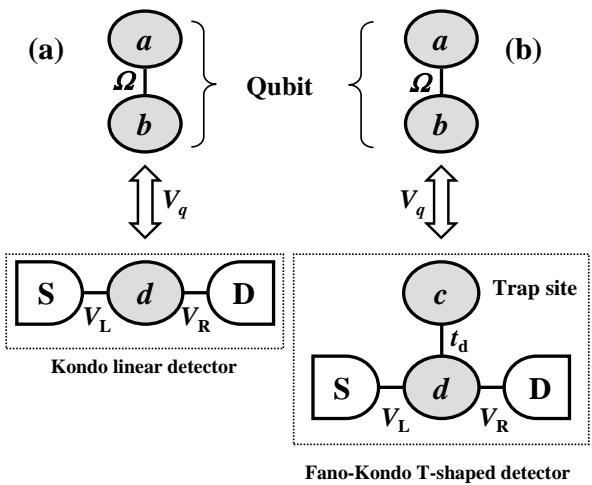}
\vspace*{6cm}
\caption{
Two types of charge qubit (two-level system) detectors. 
(a) The charge qubit detected by the Kondo effect. 
(b) The charge qubit detected by the Fano-Kondo effect.
Note that the charge qubit $``a$--$b"$ is composed of two QDs ($``a"$ and $``b"$). 
The tunneling coupling between QD $``a"$ and $``b"$ is $\Omega$.
The charge qubit is coupled to the detector part by the capacitive coupling energy $V_q$.
The detecting QDs $``d"$ are coupled to the source ``S" and the drain ``D". 
The linear-shaped ``S-$d$-``D" detector for the Kondo system in (a) is replaced by a T-shaped detector for the Fano-Kondo system in (b).
In (b), the on-site Coulomb repulsion for the trap QD $``c"$ is strong.
For the QD $``d"$, the on-site Coulomb repulsion is strong for the Kondo detector in (a) and weak for (b).
}
\label{Dots}
\end{figure}
For simplicity, and without loss of generality, we assume that all QDs have a single energy level and that 
there is a strong on-site Coulomb interaction 
in the QD $``d"$ of the Kondo detector, and the QD $``c"$ of the Fano-Kondo detector, but not for the 
QD $``d"$ of the Fano-Kondo detector.
If there is a strong on-site 
Coulomb interaction in the QD $``d"$ of the Fano-Kondo detector, the Fano resonance becomes complicated~\cite{tananishi}. 
We use a slave-boson mean-field theory (SBMFT)~\cite{Wu,Newns,tananishi,Nori} with the
help of nonequilibrium Keldysh Green functions to calculate the conductance of the detectors. 
Moreover, we assume that the interactions between the qubit and the detectors are weak
and can be decoupled into the mean-field parameters of the SBMFT.
An estimate of the effect produced by the charge fluctuations would be desirable.
However, in the SBMFT, charge fluctuations are neglected~\cite{Lopez}. 
This is a limitation of the SBMFT approach. 
We consider as following:
In the Kondo linear detector and the T-shaped QD detector setups, 
the SBMFT approach is widely used as an appropriate method to describe the system~\cite{Wu,Kang,Lopez}. 
Because our setups (Fig.~1) are based on the Kondo linear detector and T-shaped QD detector, 
as long as the coupling between the qubit and the detector is weak, 
the application of the SBMFT to our setups is a suitable starting point to treat the 
complicated electronic structure of these QD systems.

The rest of the paper is organized as follows.
In section II, we formulate the slave-boson mean-field method to calculate 
the conductance of the Kondo and the Fano-Kondo detectors. 
In section III, we show numerical results regarding the shifts of the
conductance peak and dip. 
In section IV, we use a perturbation theory to 
estimate the validity of the decoupling approximation between the qubit and the detectors. 
Sections V presents discussions and conclusions.
In the Appendix A, we summarize the derivation of the coupling constant $V_q$ 
from a network capacitance model.

\section{Formulation}
\subsection{Hamiltonian} 
As shown in Fig.~1, we study the detection of the state of a charge qubit via either the Kondo 
or Fano-Kondo effects in the detector.
The total qubit-detector Hamiltonian has three terms $H=H_{\rm det}+H_{\rm q}+H_{\rm int}$, 
where $H_{\rm det}$ describes the detector, $H_{\rm q}$ the charge qubit, 
and $H_{\rm int}$ the interaction between the charge qubit and the detector.
Here $H_{\rm q}$ is written as
\begin{equation}
H_{\rm q}=\Omega (d_a^\dagger d_b +d_b^\dagger d_a)
+\varepsilon_q ( d_a^\dagger d_a -d_b^\dagger d_b).
\label{qubit}
\end{equation}
$d_a$ and $d_b$ are electron annihilation operators of the upper QD $``a"$ and the lower QD $``b"$ 
in the charge qubit, respectively. Experimentally, $\varepsilon_q$ can be 
controlled by the gate electrode attached to the QD $``a"$ (not shown in Fig.~1). Thus, we call $\varepsilon_q$ qubit bias.
The detector Hamiltonian $H_{\rm det}$ is composed of an electrode part $H_{\rm SD}$ and a QD part 
$H_{\rm QD}$.  
Because we assume that there are strong on-site Coulomb interactions in the QD $``d"$ of the 
Kondo detector and the QD $``c"$ of the Fano-Kondo detector making double occupation of these dots impossible, 
we introduce a slave boson operator $b_d$ for the Kondo detector 
and a slave boson operator $b_c$ for the Fano-Kondo detector\cite{Newns,tananishi}.
The Kondo (K) detector Hamiltonian is  
\begin{equation}
H_{\rm det}^{(\rm K)}=H_{\rm SD}+H^{\rm (K)}_{\rm QD},
\end{equation}
and the Fano-Kondo (F) detector Hamiltonian is  
\begin{equation}
H_{\rm det}^{(\rm F)}=H_{\rm SD}+H^{\rm (F)}_{\rm QD},
\end{equation}
where
\begin{eqnarray}
H_{\rm SD}\!&\!\!=\!&\!\!\!\!\!\sum_{\alpha=L,R}\sum_{k_\alpha,s}
\{ \varepsilon_{k_\alpha}c_{k_\alpha s}^\dagger c_{k_\alpha s}
+V_\alpha (c_{k_\alpha s}^\dagger  f_{ds}
+f_{ds}^\dagger  c_{k_\alpha s}) \},
\\
H^{\rm (K)}_{\rm QD} \!&=&\!
\sum_{s}\varepsilon_{d}f_{ds}^\dagger f_{ds}
+ \lambda_{d} \left[\sum_s f_{ds}^\dagger f_{ds}
+b_{d}^\dagger b_{d}-1\right],
\\
H^{\rm (F)}_{\rm QD} \!&=&\!
\sum_{\alpha_1=c,d}\sum_{s}
\varepsilon_{\alpha_1}f_{\alpha_1s}^\dagger f_{\alpha_1s}
+ \lambda_{c} \left[\sum_s f_{cs}^\dagger f_{cs}+b_{c}^\dagger b_{c}-1\right]
\nonumber \\
&+&t_d\sum_{s}(f_{ds}^\dagger  b_c^\dagger f_{cs}
+f_{cs}^\dagger b_c f_{ds}).
\label{Hamiltonian}
\end{eqnarray}
Here $\varepsilon_{k_\alpha}$ is the energy level for the source ($\alpha=L$) and drain ($\alpha=R$) 
electrodes; $\varepsilon_{c}$ and $\varepsilon_{d}$ are energy levels for the two QDs, respectively;
$t_d$ and $V_\alpha$ are the tunneling coupling strengths 
between the trap QD $``c"$ and the detecting QD $``d"$, and 
that between QD $``d"$ and the electrodes, respectively;
$c_{k_\alpha s}$ and $f_{\alpha_1s}$ are annihilation operators of the electrodes, 
and of the QDs $(\alpha_1=c,d)$, respectively;
$s$ is the spin degree of freedom with spin degeneracy $2$;
$\lambda_{\alpha_1}$ is a Lagrange multiplier.
In the mean field theory, slave boson operators are treated as classical values such as $b_{\alpha_1} \rightarrow \langle b_{\alpha_1}\rangle$. 
We take $\langle b_{\alpha_1}\rangle$ and $\tilde{\varepsilon}_{\alpha_1} \equiv \varepsilon_{\alpha_1}+\lambda_{\alpha_1}$
as mean-field parameters that are obtained numerically by solving self-consistent equations.
The Kondo temperature is estimated as
$T_K^{\rm(K)} \sim \sqrt{\tilde{\varepsilon_d}^2 + \gamma^2 z_c^2}$ for the Kondo detector~\cite{Newns}, and 
$T_K^{\rm(F)} \sim \sqrt{\tilde{\varepsilon_c}^2 + t_d^2z_d^2}$ for the Fano-Kondo detector~\cite{Wu}, 
where $z_{\alpha_1} \equiv \langle b_{\alpha_1}^\dagger \rangle \langle b_{\alpha_1}\rangle$. 
In the numerical calculations 
shown below, we take a temperature of $T=0.02t_d<T_K^{\rm(K)}, T_K^{\rm(F)}$.

The interaction Hamiltonian $H_{\rm int}$ is derived from a capacitance network model as shown in the Appendix A~\cite{tana1,tana2}
\begin{equation}
H_{\rm int}=V_q z_{\alpha_1}\;  \sigma^z n_{\alpha_1}, \ \  (\alpha_1=c,d)
\label{int}
\end{equation}
where $n_{c}$ and $n_{d}$ are the numbers of electrons in the trap QD $``c"$, given by $n_c=\sum_s f^\dagger_{cs} f_{cs}$ 
for the Fano-Kondo case, and in the detecting QD $``d"$ is given by $n_d=\sum_s f^\dagger_{ds} f_{ds}$ for the Kondo case. 
Also, $\sigma^z$ is given by $\sigma_z= d_a^\dagger d_a -d_{b}^\dagger d_{b}$.
As shown by Eq.~(\ref{coupling1}) in the Appendix A,  $V_q \propto 1/d_D$ where $d_D$ is the distance 
between the charge qubit and the detecting QD.

We assume that the interaction between the charge qubit and the detector is weak and 
the decoupling approximation~\cite{Aono} to the interaction Hamiltonian Eq.~(\ref{int}) can be applied.
In the decoupling approximation used here, 
the electric field which the qubit senses is almost constant and we can thus 
decouple the interaction between the qubit and the detector.
The decoupling of the interaction term $H_{\rm int}$ leads to
\begin{eqnarray}
 \lefteqn{H_{\rm int}^{\rm MF}\equiv  V_q z_{\alpha_1}\left\{ \langle \sigma^z \rangle n_{\alpha_1}  +\sigma^z \langle n_{\alpha_1} \rangle 
 -\langle \sigma^z \rangle  \langle n_{\alpha_1} \rangle \right\} }  
\nonumber \\
\!\!\!\!\!\!&=& \! \!
V_q z_{\alpha_1}\Bigl\{ (\chi_{qa} -\chi_{qb})  n_{\alpha_1}\!
+[\sigma^z \!-(\chi_{qa} -\chi_{qb})]  \sum_s \chi_{\alpha_1s} \Bigr\}, \nonumber \\   
\label{decoupling}
\end{eqnarray}
where $\alpha_1=c,d$, 
$\chi_{qs} \equiv \langle d_s^\dagger d_s \rangle \label{chi_q}$, and
$\chi_{\alpha_1s}  \equiv \langle f_{\alpha_1s}^\dagger f_{\alpha_1s} \rangle$, with $\alpha_1=c,d$. 
In this decoupling approximation, $\varepsilon_{c}$,  $\varepsilon_{d}$, and $\varepsilon_{q}$ are replaced by
\begin{eqnarray}
\varepsilon_{\alpha_1}\ &\rightarrow &\ \varepsilon_{\alpha_1}'\ \equiv \ \varepsilon_{\alpha_1}
+ \lambda_{\alpha_1}+ V_q z_{\alpha_1}[\chi_{qa} -\chi_{qb}], \\
\varepsilon_{q}\ &\rightarrow &\ \varepsilon_{q}' \ \equiv \ \varepsilon_q + V_q z_{\alpha_1}\sum_s \chi_{\alpha_1s}.
\label{replace}
\end{eqnarray}

\subsection{Green functions} 
Charge qubit detection in our system is carried out by the measurement of the current of the detector.
The current through each detector is calculated using the non-equilibrium Green functions as~\cite{Wingreen,tananishi}
\begin{eqnarray}
J&=&\frac{ie}{\hbar}\sum_{k_L,s} [V_L^* \langle c_{kLs}^\dagger f_{ds} \rangle
-V_L \langle f_{ds}^\dagger c_{kLs} \rangle ]
\nonumber \\
&=&\frac{2e}{\hbar}\sum_{k_L,s} {\rm Re} \left[ V_L^* G_{dk}^<(t,t) \right] 
\nonumber \\
&=&\frac{2e}{\hbar} {\rm Re}  \sum_{k_L,s} \int d \omega \left[ V_L^* G_{dk}^<(\omega) \right], 
\end{eqnarray}
where $G_{dk}^<(t,t)\equiv \langle c_{kLs}^\dagger(t) f_{ds} (t)\rangle$, and 
\begin{equation}
G_{dk}^<(\omega)=V_L[g_{kL}^r(\omega) G_{dd}^<(\omega) + g_{kL}^<(\omega) G_{dd}^a(\omega)],
\end{equation}
with $g_{kL}^r(\omega)\equiv(\omega-\varepsilon_{kL}+i\delta)^{-1}$ ($\delta$ is an infinitesimal quantity) and  
$g_{kL}^<(\omega)\equiv 2\pi \delta(\omega-\varepsilon_{kL}) f_L(\omega)$. 
Here $f_L (\omega)\equiv \{ \exp [(\omega+eV_{\rm bias}-E_F)/(k_BT)]+1\}^{-1}$ and
$f_R (\omega)\equiv \{ \exp [(\omega-E_F)/(k_BT)]+1\}^{-1}$ are the Fermi distribution functions of the electrodes 
when there is a finite bias voltage $V_{\rm bias}$ between the
two electrodes ($k_B$ is the Boltzmann constant). 

Let us first consider the Green function formulation for the Fano-Kondo system. 
With the decoupling given in Eq.~(\ref{decoupling}), the Green functions remain the same as those without interactions 
between the charge qubit and the detector, by changing the replacements Eq.~(\ref{replace}).
Using the equation of motion method, the advanced Green function $G^{\rm a}$ for the detecting QD $``d"$ 
of the Fano-Kondo detector is obtained as
\begin{equation}
G_{dd}^{\rm (F) a} =
\frac{\omega-\varepsilon_c' -i \delta}{(\omega-\varepsilon_d-i\gamma)(\omega-\varepsilon_c' -i \delta) -|\tilde{t}_d|^2},  
\label{FG}
\end{equation}
where $\tilde{t}_d=t_d\langle b_c \rangle$.
The $G_{dd}^{({\rm F})<} (t) \equiv -i\langle f_d^\dagger(t)f_d(t) \rangle $ can then be calculated from $G_{dd}^<=G_{dd}^{\rm r} \Sigma_{dd}^< G_{dd}^{\rm a}$ with
\begin{equation}
\Sigma_{dd}^<(\omega) 
=i[\Gamma_L f_L(\omega) +\Gamma_R f_R(\omega)],
\end{equation}
where $\Gamma_\alpha\equiv 2\pi \rho_\alpha (E_F)|V_\alpha |^2$ is the tunneling rate 
between the $\alpha$ electrode ($\alpha=L,R$) and the detecting QD $``d"$, with a density of states (DOS) $\rho_\alpha (E_F)$ 
for each electrode at the Fermi energy $E_F$. 
Thus
\begin{equation}
G_{dd}^{({\rm F})<}= \frac{ i\chi(\omega)}{C_{00}} (\omega-\varepsilon_c')^2,
\end{equation}
where 
\begin{eqnarray}
\chi (\omega)\!& \equiv&\!\Gamma_L f_L(\omega) +\Gamma_R f_R(\omega), \\
C_{00}\!&\equiv\!&\! [(\omega-\varepsilon_c')(\omega-\varepsilon_d)-|\tilde{t}_d|^2]^2 +\gamma^2 (\omega-\varepsilon_c')^2. 
\end{eqnarray}
Similarly, we find
\begin{eqnarray}
G_{cd}^{({\rm F})<} (\omega) 
&=& \frac{i\chi (\omega)(\omega-\varepsilon_c')}{C_{00}}, \\ 
G_{cc}^{({\rm F})<}(\omega)&=& \frac{i\chi(\omega) |\tilde{t}_d|^2}{C_{00}},
\end{eqnarray}
with $\gamma=(\Gamma_L + \Gamma_R)/2$.
We also define $\Gamma\equiv 2 \Gamma_L\Gamma_R/(\Gamma_L+\Gamma_R)$. 
For simplicity, we assume $\Gamma_L=\Gamma_R$.

For the Kondo linear detector shown in Fig.~1(a), 
the Green functions are similarly obtained by using the equation of motion method
\begin{eqnarray}
G_{dd}^{({\rm K})<}&=& \frac{ iz_d\chi(\omega)}{(\omega-\varepsilon_d')^2+\gamma^2z_d^2}, \\
G_{dd}^{\rm(K)a} &=&
\frac{1}{\omega-\varepsilon_d'-i\gamma z_d}. 
\label{KG}
\end{eqnarray}
The charge qubit Green functions are expressed as
\begin{eqnarray}
G_{aa}(\omega) &=& \frac{\omega+\varepsilon_{q}'}{(\omega-\Delta)(\omega + \Delta)}, \\
G_{bb}(\omega) &=& \frac{\omega-\varepsilon_{q}'}{(\omega-\Delta)(\omega + \Delta)},  \\
G_{ab}(\omega) &=& G_{ba}(\omega) = \frac{\Omega}{\omega^2 -\varepsilon_{q}'^2-\Omega}.
\end{eqnarray}
where $\Delta\equiv \sqrt{\varepsilon_{q}'^2+ \Omega^2}$.

Thus, the current for the Fano-Kondo detector can be expressed as
\begin{equation}
J=\frac{e}{\hbar} \int \frac{d\omega}{\pi} \frac{z_d^2 \Gamma_L\Gamma_R(\omega-\varepsilon_c')^2}{C_{00}}
\left[f_L(\omega)-f_R(\omega)\right],
\label{current_K}
\end{equation}
and the current for the Kondo detector is given by
\begin{equation}
J=\frac{e}{\hbar} \int \frac{d\omega}{\pi} \frac{z_d^2 \Gamma_L\Gamma_R}{(\omega -\varepsilon_d')^2 +\gamma^2 z_d^2}
\left[f_L(\omega)-f_R(\omega)\right].
\label{current_FK}
\end{equation}
In Sec. III, we show numerical results of conductance $G\equiv dJ/dV_{\rm bias}$ at $V_{\rm bias}=0$,  
and discuss the transport properties of the two detector.

\begin{figure}
\includegraphics[width=3.3in]{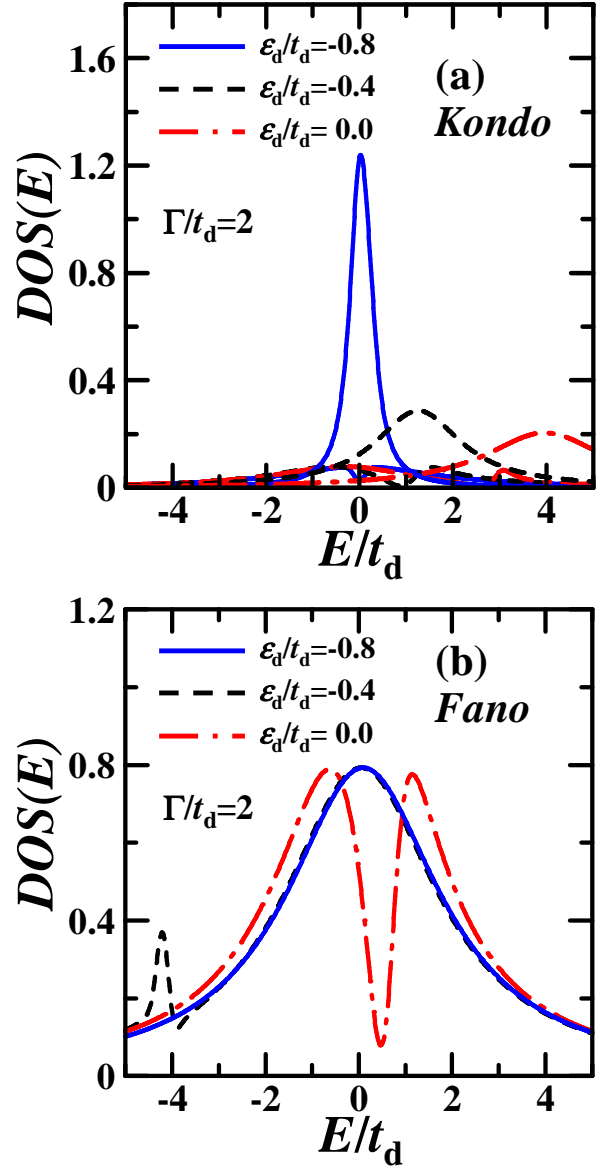}
\caption{(Color online) Density of states (DOS) of the detecting QD $``d"$ for (a) the Kondo detector and (b) the Fano-Kondo detector, 
using $\Omega/t_d=1$, $V_q/t_d=0.5$, $\varepsilon_q/t_d=-0.05$, and $T/t_d=0.02$.}
\label{FigDOS}
\end{figure}

\subsection{Self-consistent equations} 
The detector current is calculated self-consistently: while the qubit state influences the detector QD energy level 
and thus the current through it, the qubit state itself is also affected by the detector QD occupation 
through capacitive coupling, as described by Eq.~(\ref{decoupling}).  
Here we derive the self-consistent equations.
The DOS of qubits are derived from the qubit Green function 
$\rho_{l}=-\frac{1}{\pi} {\rm Im}G_{ll}(\omega+i\delta)$ 
($l=a,b$).
Then, the average electron occupancy $\chi_{ql }$ of the two QDs of the qubit is expressed by
\begin{equation}
\chi_{ql }=\int_{-D}^{D} \!\!d\omega f(\omega) \rho_{l}(\omega) 
=\frac{1}{2} \left( 1 +p_l \frac{\varepsilon_q'}{\Delta} \tanh \frac{\beta\Delta}{2} \right), 
\end{equation}
where $p_a=1$, $p_b=-1$, and $f(\omega)\equiv \{\exp [(\omega-E_F)/(k_BT)]+1\}^{-1}$. 
Using Eq.~(\ref{replace}), we have the self-consistent equations for the Fano-Kondo case:
\begin{eqnarray}
& & \varepsilon_{c}'= \varepsilon_{c}+\lambda_{c}+V_q z_c \frac{\varepsilon_q'}{\Delta} \tanh \frac{\beta\Delta}{2},  \label{sq3}\\
& & \varepsilon_{q}' =\varepsilon_q + V_q z_c [1-  z_c], \label{sq4} \\
& & \int \frac{d\omega}{\pi} \frac{(\omega-\varepsilon_c')|t_d|^2 }{C_{00}} \; \chi (\omega) +\lambda_c+
\frac{\partial \varepsilon_c' }{\partial z_c} =0,
\\
& &\int \frac{d\omega}{\pi} \frac{z_c|t_d|^2}{C_{00}}\; \chi(\omega) +z_c-1 =0, 
\label{sq5}
\end{eqnarray}
where $z_c = |\langle b_c \rangle|^2$ and $\beta^{-1}=k_B T$. 
From Eq.~(\ref{sq4}), we can see that the energy shift $\varepsilon_q'-\varepsilon_q$
of the charge qubit is related to the electron occupancy $1-z_c=1-|\langle b_c \rangle|^2$ of the trap site $``c"$,   
and its magnitude is proportional to the coupling strength $V_q$.
In particular,  the qubit energy shifts as a function of $V_q$ due to {\it back-action}.

For the Kondo detector interacting with the charge qubit, the self-consistent equations are
\begin{eqnarray}
& & \varepsilon_{d}'= \varepsilon_{d}+\lambda_{d}+V_q z_d \frac{\varepsilon_q'}{\Delta} \tanh \frac{\beta\Delta}{2},  \label{sq3a}\\
& & \varepsilon_{q}' =\varepsilon_q + V_q z_d[1-  z_d], \label{sq4a} \\
& & \int \frac{d\omega}{\pi} \frac{(\omega-\varepsilon_d')}{(\omega -\varepsilon_d')^2 +\gamma^2 z_d^2} \chi (\omega) +\lambda_d+
\frac{\partial \varepsilon_d' }{\partial z_d} =0,
\\
& &\int \frac{d\omega}{\pi} \frac{z_d}{(\omega -\varepsilon_d')^2 +\gamma^2 z_d^2}\chi(\omega) +z_d-1 =0.
\label{sq6}
\end{eqnarray}

\section{Numerical results}
Here we show numerical results focusing on the shift of the conductance {\it peak} 
in the Kondo effect and the shift of the conductance {\it dip} of the Fano-Kondo effect. 
Although $t_d$ appears only in the Fano-Kondo detector, we measure all energies in units of $t_d$, to better  
compare the Kondo detector with the Fano-Kondo detector.
For the Fano-Kondo detector, when $\Gamma \gg t_d$, the electron tunneling between the QD $``c"$ and 
the QD $``d"$ cannot be easily observed because the current flow to and from the two electrodes is too fast, 
so that it drowns out the effects of the electron tunneling between QDs $``c"$ an $``d"$. 
Thus, as shown in Ref.~\onlinecite{tananishi}, 
we use the QD-electrode tunneling rate $\Gamma$ to characterize the detection speed.  
Specifically, we denote the case of $\Gamma/t_d=2$ as a fast detector, and $\Gamma/t_d=0.04$ as a slow detector

\subsection{Conductance}
Figure~\ref{FigDOS} shows numerical results of the DOS of the detecting QDs $``d"$. 
The DOS $\rho_{\rm det}(\omega)$ of the detector QD is derived from
$\rho_{\rm det}(\omega)\equiv -{\rm Im}G_{dd}^r (\omega)/\pi$. 
For the Kondo case in (a) there is a single peak, and for the Fano-Kondo case in (b) we can see the Fano 
asymmetric line shape. 
Figure~\ref{FigG} shows the conductance of the Kondo detector [(a) and (c)] and the Fano-Kondo detector 
[(b) and (d)], as a function of the detector QD energy levels $\varepsilon_{d}$ of the Kondo detector [(a) and (c)] 
and $\varepsilon_{c}$ the Fano-Kondo detector [(b) and (d)], respectively.
We can see clear peaks for the Kondo detector [(a) and (c)], and clear dips for the Fano-Kondo detector 
[(b) and (d)], as in Ref.~\onlinecite{Wingreen,Hofstetter,Sasaki1,Sasaki3,Kogan,Kobayashi,Dagotto,Wu,Kang,Ueda,Gores,Sato,Rushforth,Sasaki2}.
These peaks and dips are maximized when the coherence 
between the discrete energy state and the continuum states is largest, we thus denote corresponding
energies $\varepsilon_d^{\rm (peak)}$ and $\varepsilon_c^{\rm (dip)}$ as {\it coherent extrema}.
For the Kondo detector, because of Eq.~(\ref{KG}), as the detector speed $\Gamma$ increases, 
the width of the peak also increases. However, for the Fano-Kondo detector, 
because of Eq.~(\ref{FG}), as the detector speed increases, the width of the dip decreases.
For both detectors, the shifts of the conductance peaks and dips are observed, when $\varepsilon_q$ is changed.
Below we investigate the shift of the coherent extrema $\varepsilon_d^{(\rm peak)}$ and $\varepsilon_c^{(\rm dip)}$ in more detail.
\begin{widetext}
\begin{center}
\begin{figure}
\includegraphics[width=14cm]{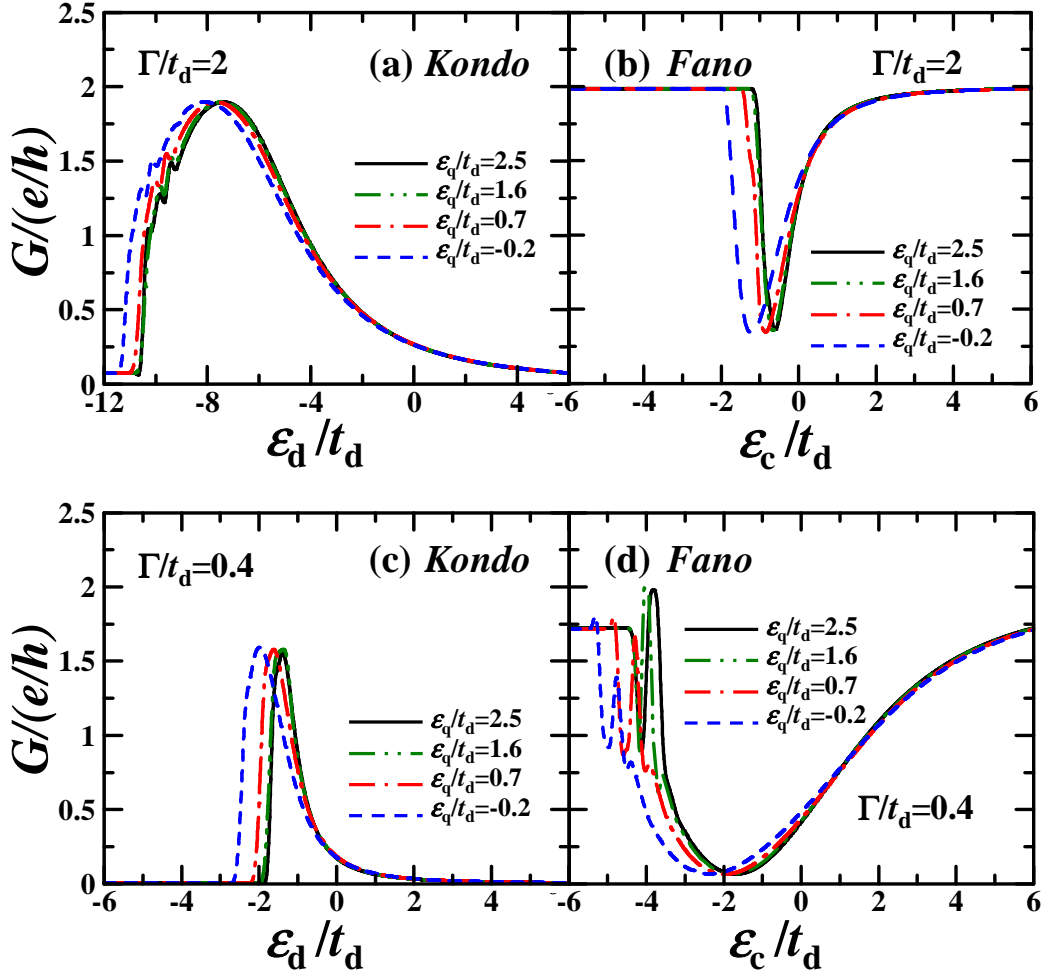}
\vspace*{12.8cm}
\caption{(Color online) 
Numerical results for the conductance $G$ (in units of $e/h$) 
for detectors as a function of the QD energies ($\varepsilon_d$ for the detecting QD $``d"$ of the Kondo detector, 
and $\varepsilon_c$ for the charge trap $``c"$ of the Fano-Kondo detector) for $\Omega/t_d=1$, $V_q/t_d=0.5$ and 
temperature $T/t_d=0.02$.
(a) Fast Kondo detector :$\Gamma/t_d=2$. 
(b) Fast Fano-Kondo detector :$\Gamma/t_d=2$. 
(c) Slow Kondo detector :$\Gamma/t_d=0.4$. 
(d) Slow Fano-Kondo detector :$\Gamma/t_d=0.4$. 
The peak positions for the Kondo detector and the dip positions for the Fano-Kondo effect 
are shifted by the qubit bias $\varepsilon_q$.
}
\label{FigG}
\end{figure}

\begin{figure}
\includegraphics[width=14cm]{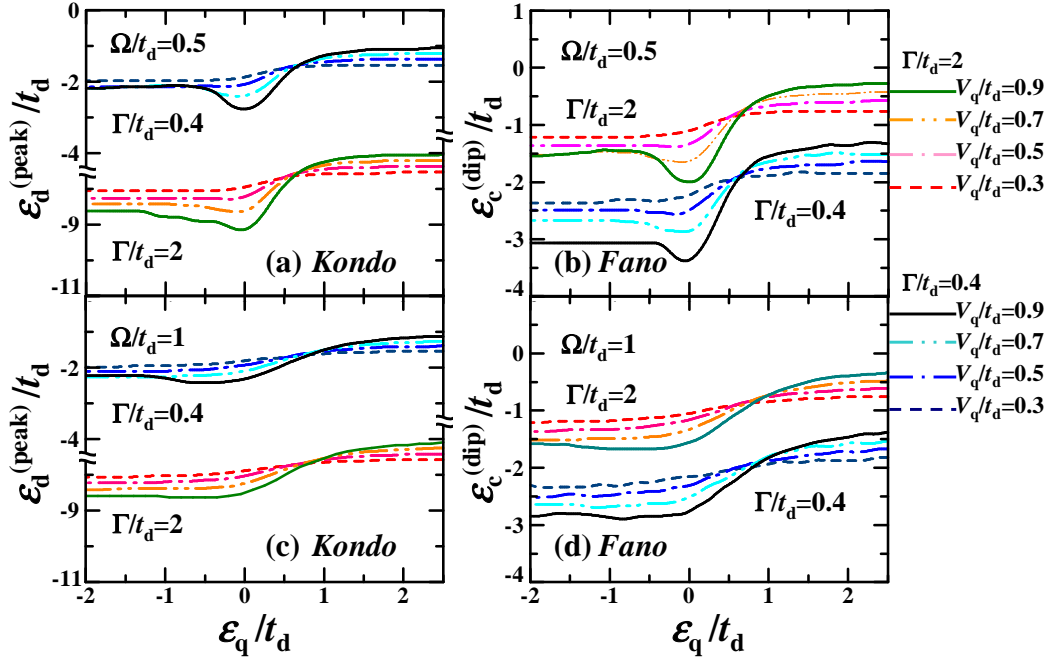}
\caption{(Color online)
The coherent extrema $\varepsilon_d^{\rm (peak)}$ (conductance peak) 
and  $\varepsilon_c^{\rm (dip)}$ (conductance dip), as a function of  the qubit bias $\varepsilon_q$.
The conductance peak of the Kondo detector for the $\Omega/t_d=0.5$ qubit (a), and 
the $\Omega/t_d=1$ qubit (c). The conductance dip of the Fano-Kondo detector 
for the $\Omega/t_d=0.5$ qubit (b), and 
the $\Omega/t_d=1$ qubit (d). 
The $\varepsilon_d^{\rm (peak)}$  and $\varepsilon_c^{\rm (dip)}$  
are smallest around the optimal point $\varepsilon_q' =\varepsilon_q \approx 0$.
}
\label{FigDiffE}
\end{figure}
\end{center}

\begin{figure}
\includegraphics[width=14cm]{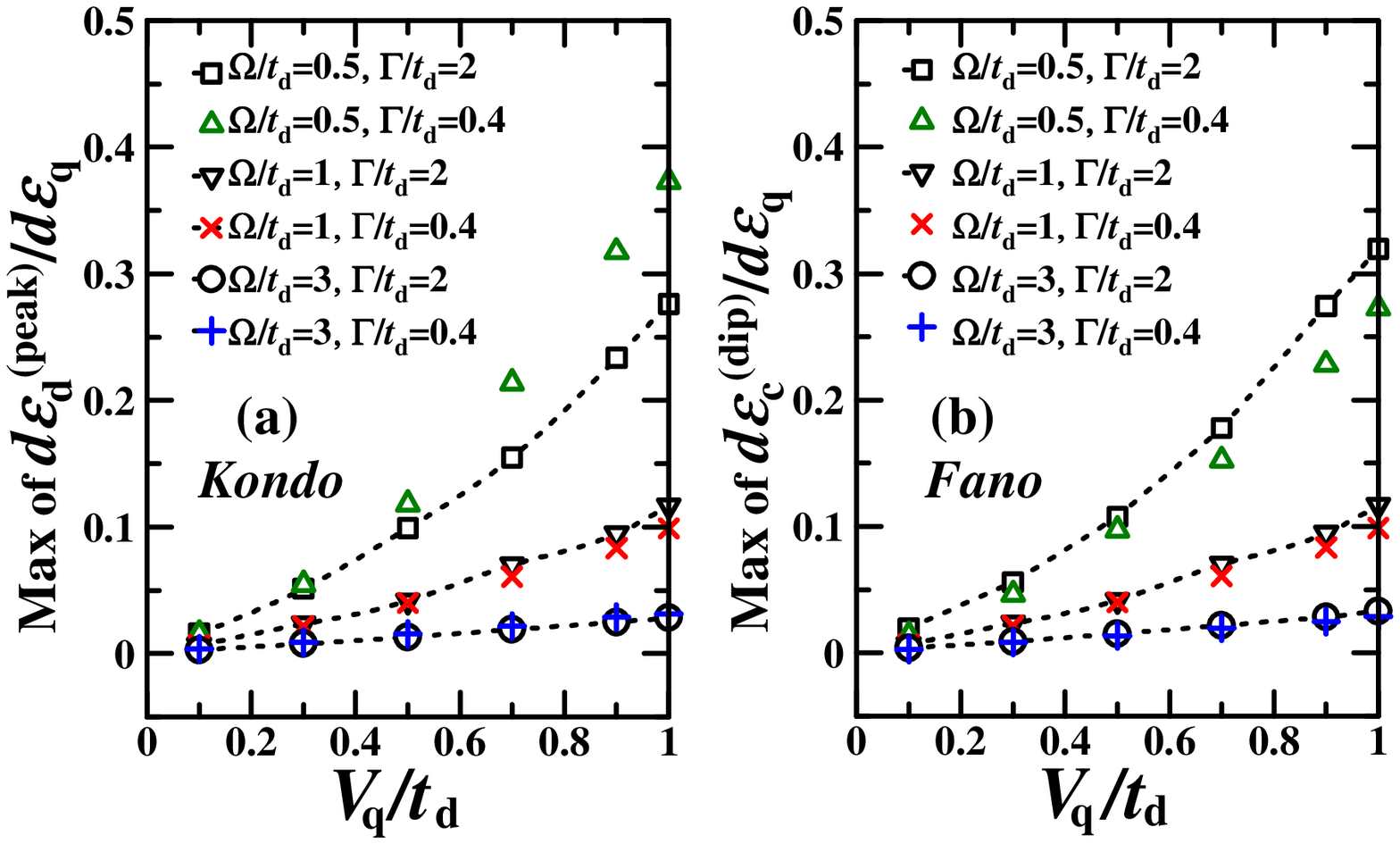}
\caption{(Color online)
Maximum values of  $d\varepsilon_d^{\rm (peak)}/d\varepsilon_q$ and $d\varepsilon_c^{\rm (dip)}/d\varepsilon_q$ plotted 
as a function of the qubit-detector coupling $V_q$ for both 
fast ($\Gamma/t_d=2$) and slow ($\Gamma/t_d=0.4$) detectors 
for  (a) the Kondo and (b) the Fano-Kondo detectors.
It can be seen that the maximum values do {\it not} vary with the speed $\Gamma$ of the detectors.
It can also be seen that  $d\varepsilon_d^{\rm (peak)}/d\varepsilon_q \propto V_q/\Omega$
and  $d\varepsilon_c^{\rm (dip)}/d\varepsilon_q \propto V_q/\Omega$.
}
\label{FigPeak}
\end{figure}
\end{widetext}

Figure~\ref{FigDiffE} plots the coherent extrema $\varepsilon_d^{\rm (peak)}$ 
and  $\varepsilon_c^{\rm (dip)}$, as a function of  the qubit bias $\varepsilon_q$ [Eq.~(\ref{qubit})].
As the qubit bias $\varepsilon_q$ increases, the distribution of the excess charge 
in the qubit approaches to the detector QDs, resulting in raising the energy
of QD $``d"$ of the Kondo detector and that of QD $``c"$ of the Fano-Kondo detector.
Finally, the increase of the QD energies are saturated because of the balance of the charge distribution.
Figure~\ref{FigDiffE} reflects this fact and shows that 
$\varepsilon_d^{\rm (peak)}$ and $\varepsilon_c^{\rm (dip)}$ increase as $\varepsilon_q$ increase. 

Because $z_{\alpha_1}=0$ ($\alpha_1=c,d$) is satisfied at the coherent extrema, we have the relation $\varepsilon_q'=\varepsilon_q$ 
from Eq.~(\ref{sq4}) and Eq.~(\ref{sq4a}).
Then, it can be observed that the minimum of $\varepsilon_d^{\rm (peak)}$ and $\varepsilon_c^{\rm (dip)}$ exist 
around the $\varepsilon_q'\sim 0$ region in Fig.~\ref{FigDiffE}.  
At $\varepsilon_q' \sim 0$, the energy splitting $\sqrt{\Omega^2+\varepsilon_q'^2}$ 
between the two eigenenergies of the qubit is smallest, and the qubit energy splitting is a quadratic function of the qubit bias. 
Thus, qubit state is insensitive 
to charge noises that lead to qubit dephasing, and this zero bias point corresponds to an optimal point, 
in analogy to other similar cases~\cite{Vion,TanaHu,Li}. 

The third terms of Eq.~(\ref{sq3}) and Eq.~(\ref{sq3a}) decrease $\varepsilon_d^{\rm (peak)}$  
and  $\varepsilon_c^{\rm (dip)}$ when $\varepsilon_q'<0$ and increases them when $\varepsilon_q'>0$  
($\beta \Delta \gg 1$), resulting in the minimum structure of Fig.~\ref{FigDiffE} at the optimal point $\varepsilon_q'=0$. 
In addition, because the third terms of Eq.~(\ref{sq3}) and Eq.~(\ref{sq3a}) become larger as $\Omega$ becomes smaller,  
Fig.~\ref{FigDiffE} (a,b) are considered to show clearer minimum structures than Fig.~\ref{FigDiffE} (c,d). 
We can also observe that the magnitude of the minimum is proportional to the coupling strength $V_q$.
This is also the reason that the minimum in Fig.~\ref{FigDiffE} are caused by the third terms of Eq.~(\ref{sq3}) and Eq.~(\ref{sq3a}).
From Eq.~(\ref{sq3}) and Eq.~(\ref{sq3a}), at the optimal point ($\varepsilon_q'=0$, thus, $\Delta =\Omega$) of the coherent extrema, 
we obtain 
\begin{equation}
\frac{d\varepsilon_{\alpha_1}}{d{\varepsilon_q}} \approx\frac{d\lambda_{\alpha_1}}{d\varepsilon_q}.
\label{diffa}
\end{equation} 
($\alpha_1=c,d$). 
The peaks of Fig.~\ref{FigDiffE} correspond to ${d\varepsilon_{\alpha_1}}/{d{\varepsilon_q}}=0$ and  
Eq.~(\ref{diffa}) shows that $d\lambda_{\alpha_1}/d\varepsilon_q=0$ 
at the minimum values of the coherent extrema. 

As mentioned above, the $\varepsilon_d^{\rm (peak)}$ and $\varepsilon_c^{\rm (dip)}$ increase  when $\varepsilon_q$ increases, 
and they are finally saturated after they have their minimum regarding the optimal points. Thus, their derivatives, 
$d\varepsilon_d^{\rm (peak)}/d\varepsilon_q$ and $d\varepsilon_c^{\rm (dip)}/d\varepsilon_q$, 
are considered to have their maximum values around the middle points between the minimum of the optimal points and 
the small $\varepsilon_q$ region. 
Figure~\ref{FigPeak} plots the maximum values of  $d\varepsilon_d^{\rm (peak)}/d\varepsilon_q$ 
and $d\varepsilon_c^{\rm (dip)}/d\varepsilon_q$ as a function of the qubit-detector coupling $V_q$.
It can be seen that: 
(i) the maximum values of  $d\varepsilon_d^{\rm (peak)}/d\varepsilon_q$ 
and $d\varepsilon_c^{\rm (dip)}/d\varepsilon_q$ do not depend on the speed $\Gamma$ of the detectors, and 
(ii) there is a relationship between the peaks and $V_q/\Omega$ such as
\begin{eqnarray} 
{\rm Max} \frac{d\varepsilon_d^{\rm (peak)}}{d{\varepsilon_q}}&\propto& \frac{V_q}{\Omega} \label{maxd},\\
{\rm Max} \frac{d\varepsilon_c^{\rm (dip)}}{d{\varepsilon_q}}&\propto& \frac{V_q}{\Omega} \label{maxc},
\end{eqnarray}
when $\Omega/t_d >1$. 
These weak dependences of the maximum values on the speed $\Gamma$ of the detectors are considered to be 
because of the sharp response 
of the Kondo and Fano-Kondo effects at their coherent extrema. 
Because all quantities are numerically derived from the self-consistent equations, 
Eq.~(\ref{maxd}) and Eq.~(\ref{maxc}) cannot be derived analytically, and 
these results are obtained numerically.
In principle, $V_q$ can be calculated from the structure of the system by using the capacitance network model, 
as shown in the Appendix A. Thus, in experiments, if we can prepare several samples with the different distances between the detector and the qubit, 
we can estimate the tunneling coupling $\Omega$ for the charge qubit by using the relations Eq.~(\ref{maxd}) and Eq.~(\ref{maxc}).

Therefore, Fig.~\ref{FigDiffE} and \ref{FigPeak} indicate that 
by finding the minimum of $\varepsilon_d^{\rm (peak)}$ and $\varepsilon_c^{\rm (dip)}$, 
we can find the optimal point ($\varepsilon_q'=0$) of the qubit, and 
by analyzing the coefficients of  $d\varepsilon_d^{\rm (peak)}/d\varepsilon_q$ and $d\varepsilon_c^{\rm (dip)}/d\varepsilon_q$ 
as a function of $V_q$, we can infer the tunneling coupling $\Omega$ for the charge qubit.

Here, we check whether the temperature $T=0.002t_d$ is 
below $T_K\sim \sqrt{\tilde{\epsilon}_\alpha^2 + t_\alpha^2 z_\alpha^2}$ or not.
In our calculations, $\tilde{\epsilon}_\alpha \sim \epsilon_\alpha$ ($\alpha=c,d$).  
As can be seen from Fig.~\ref{FigG}, Kondo peaks and Fano dips are observed for $|\epsilon_\alpha| > 0.5$. 
In addition, when there are no Kondo peaks or Fano dips, $z_\alpha \sim 1$.  
Thus, in both the Kondo region and the non-Kondo region, $T_K> T=0.002t_d$ is always held. 

\subsection{Back-action}
As we have seen, both the Kondo detector and the Fano-Kondo detector 
have similar capabilities to detect the tunneling $\Omega$ and the qubit bias $\varepsilon_q'$. Here 
we consider the effect of measurement (back-action on the qubit) and the noise characteristics of the two types of detectors.
Figure \ref{FigFF}(a,b) show how  $\varepsilon_q'$ is affected by the detectors.
The change of qubit energies clearly depends on the coherent extrema of the Kondo peak and the Fano-Kondo dip.
Although figures are not shown, 
the changes of $\varepsilon_q'$ for $\Gamma/t_d=2$ of the Kondo detector 
and that for $\Gamma/t_d=0.4$ of the Fano-Kondo 
detector are larger than those in Fig.~\ref{FigFF}(a,b), respectively. 
Thus the slow Kondo detector and the fast Fano-Kondo detector 
are better from the viewpoint of back-action.

The ratio of the shot noise $S_I$
and the full Poisson noise $2eI$, $F\equiv S_I /(2eI )$, is called the Fano
factor. It indicates important noise properties with regard to
the quantum correlations~\cite{Kobayashi}. 
Smaller $F$ is better because smaller $F$ means less noise of the detection.
Similarly to the result of Ref.~\onlinecite{fanofactor},
the Fano factor $F$ at zero bias and zero temperature is given by $1-\mathcal{T}(E_F)$, 
where $\mathcal{T}(E_F)$ is a transmission probability expressed by
\begin{equation}
\mathcal{T}(\omega)\equiv \frac{2\Gamma_L\Gamma_R}{\Gamma_L+\Gamma_R}
\pi \rho_{\rm det}(\omega).
\end{equation} 
($\rho_{\rm det}(\omega)$ is the DOS of the detector QD, as mentioned above).
This means that the larger $\mathcal{T}(\omega)$ is better from the viewpoint of the noise reduction.
As can be inferred from Fig.~\ref{FigG}(a,c), $\mathcal{T}(\omega)$ for $\Gamma/t_d=2$ of the Kondo detector 
is larger than that of $\Gamma/t_d=0.4$ of the Kondo detector. 
This means that, in the case of the Kondo detector,  $F$ for $\Gamma/t_d=2$  is smaller than 
that for $\Gamma/t_d=0.4$ (Fig.~\ref{FigFF}(c)).
Similarly, in the Fano-Kondo detector, 
$F$ for $\Gamma/t_d=0.4$ is smaller that that for $\Gamma/t_d=2$ (Fig.~\ref{FigFF}(d)). 
Thus, the fast Kondo detector and the slow Fano-Kondo detector are better from viewpoint of the noise reduction.
Therefore, the magnitude of the back-action and the efficiency of the detector have a tradeoff relationship.
More advanced analysis such as Ref.~\onlinecite{Koch} should be considered as a future problem.

\begin{widetext}
\begin{center}
\begin{figure}
\includegraphics[width=13cm]{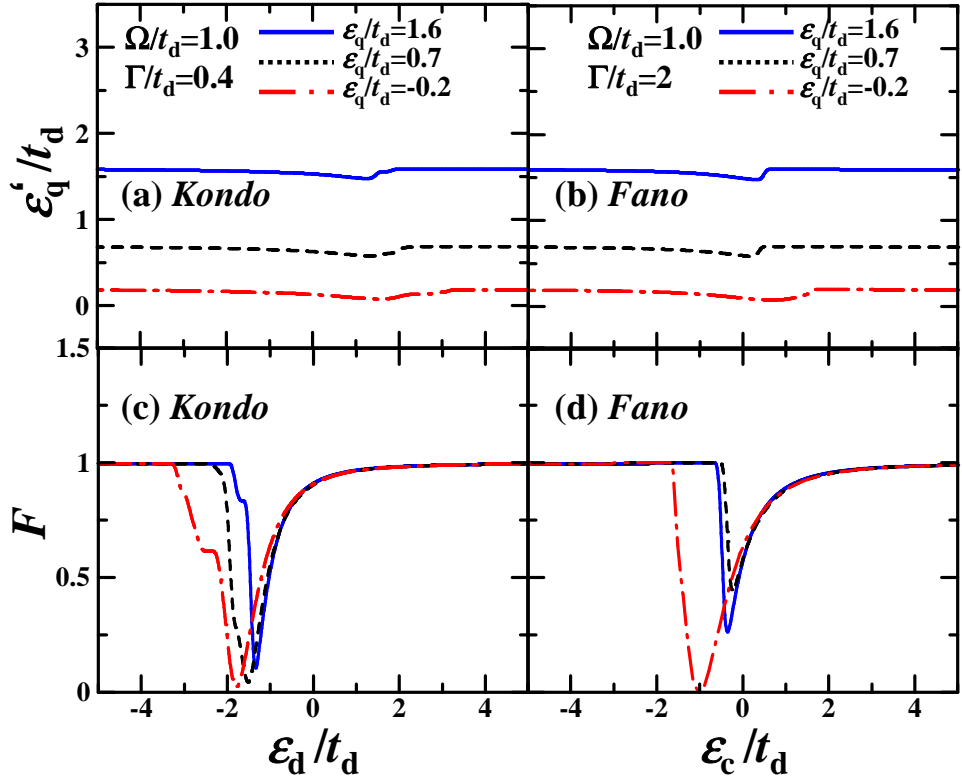}
\caption{(Color online)
The qubit bias $\varepsilon_q'$ of (a) the Kondo detector for $\Gamma/t_d=0.4$
and (b) the Fano-Kondo detector for  $\Gamma/t_d=2$. 
Fano Factor $F$ of (c) the Kondo detector for $\Gamma/t_d=0.4$ and 
(d) that of the Fano-Kondo detector for $\Gamma/t_d=2$.  
Here, $\Omega/t_d=0.5$, $V_q/t_d=0.5$, at zero temperature.}
\label{FigFF}
\end{figure}
\end{center}
\end{widetext}

\section{Perturbation theory}
\label{perturbation}
A crucial assumption that allows our calculations mentioned above to proceed is the
decoupling approximation as stated in Eq.~(\ref{decoupling}).  
Here we investigate the validity of this approximation by using a simple model in which the charge qubit is 
capacitively coupled to a QD connected to a Fermi sea.
In Ref.~\onlinecite{Makhlin}, $\Gamma$ is related to the measurement speed of the system.
Here we use $\Gamma^{-1}$, the tunneling time between the central QD and the leads, to represent 
the time scale of the detector and its temporal sensitivity.
The perturbation Hamiltonian is
\begin{eqnarray}
H_1&\equiv& H_{\rm int} -H_{\rm int}^{\rm MF} 
%
= V_q \biggl\{ \sigma_z n_{\alpha_1}
-(\chi_{qa} -\chi_{qb}) n_{\alpha_1}  
\nonumber \\
&-& \biggl(\sum_s \chi_{\alpha_1 s}\biggl) \sigma_z  
 +(\chi_{qa} -\chi_{qb}) \sum_s \chi_{\alpha_1s} \biggl\},
\end{eqnarray}
where $\alpha_1=c,d$.
Because the qubit Hamiltonian Eq.~(1) includes $\sigma_x$ and $\sigma_z$, $H_1$ can flip the 
qubit state between $|0\rangle$ and $|1\rangle$.
We apply the golden rule and calculate the transition probability 
starting from the initial qubit state $|0\rangle$. 
The transition probability $P(\Delta)$ is given by
\begin{eqnarray}
\lefteqn{ P(\Delta)=\frac{1}{\hbar^2} \sum_i \rho_i \int_{-\infty}^{\infty} \! \! \!
dt \: \langle i | H_1^\dagger (t) H_1(0) | i\rangle}
\nonumber \\
&=& \frac{V_q^2}{\hbar^2}\int_{-\infty}^{\infty}\! \! \!
dt\sum_i \rho_i \, \langle i | n_{\alpha_1}(t) n_{\alpha_1}(0) | i \rangle \; e^{i\Delta t}, 
\end{eqnarray}
where $i$ labels the eigenstates of the environment (electrodes), and $\rho_i =\exp \left( {-\beta \varepsilon_i} /Z_0\right)$, 
with an equilibrium environment partition function $Z_0$.
At zero temperature, we can decouple 
$\langle i | n_{\alpha_1}(t) n_{\alpha_1}(0) | i \rangle$  into $\langle  f_{\alpha_1}^\dagger (t) f_{\alpha_1}(0) \rangle$
and $\langle  f_{\alpha_1} (t) f_{\alpha_1}^\dagger(0)  \rangle $ using the Bloch-De Dominicis theorem~\cite{Abe}.
For the Fano-Kondo case, 
\begin{eqnarray}
\langle  f_c^\dagger (t) f_c(0) \rangle 
&=&|\tilde{t}_d|^2\int \frac{d\omega}{2\pi} \frac{ \chi (\omega)}{C_{00}}e^{i\omega t},
\\
\langle  f_c (t) f_c^\dagger(0)  \rangle 
&=&|\tilde{t}_d|^2\int \frac{d\omega}{2\pi} \frac{ \nu (\omega)}{C_{00}}e^{-i\omega t},
\end{eqnarray}
where $\nu(\omega)\equiv \sum_{\alpha=L,R}\Gamma_\alpha [1-f_\alpha (\omega)]$.
Defining the lifetime of the mean-field approximation by $1/\tau \equiv P(\Delta)$, as discussed in Ref.~\onlinecite{Leggett},
we obtain, for $\Delta \ll \gamma$
\begin{equation}
\frac{1}{\tau}\; \approx\; \frac{16 \, \Gamma^2 \, V_q^2 \, \Delta}{t_d^4z_c^2}, 
\end{equation}
and for $\Delta \gg \gamma$
\begin{equation}
\frac{1}{\tau} \; \approx \; \frac{ 4 \Gamma^2 \, V_q^2 \, t_d^4\, z_c^2 }{(a_+-a_-)\Delta^5} \log \left|\frac{a_-}{a_+}\right|,
\end{equation}
where 
\begin{equation}
a_\pm \equiv \frac{1}{2} \left(-\gamma^2+t_d^2 z_c \pm\sqrt{\gamma^2(\gamma^2-2t_d^2 z_c)}\right).
\end{equation}
For the Kondo case, we obtain for $\Delta \ll \gamma$
\begin{equation}
\frac{1}{\tau}\; \approx\; \frac{V_q^2 \, \Delta}{\gamma^2z_d^2}, 
\end{equation}
and for $\Delta \gg \gamma$
\begin{equation}
\frac{1}{\tau} \; \approx \; \frac{ 8 \Gamma^2 \, V_q^2 \,  z_d^2 }{\Delta^3} \log \left(\frac{\Delta}{\gamma z_d}\right).
\end{equation}
We estimate this lifetime for the case of $\Delta \ll \gamma$ by referring to the experimental values in Ref.~\onlinecite{Sato}. 
The intrinsic measurement time $t_{\rm m}$ can be estimated from $t_{\rm m} \approx \hbar/\Gamma$. 
When using $t_d=0.5$~meV, $\Gamma=\gamma=0.2$~meV, $V_q=0.01$~meV, and $\Delta=0.01$~meV,
we obtain $\tau\sim 64$~ns for the Fano-Kondo case, 
and we obtain $\tau\sim 26$~ns for the Kondo case. 
Then, $t_{\rm m} \approx \hbar/\Gamma \sim 0.0033 $~ps and $t_{\rm m} \ll \tau$ is held.
Thus,  in this region, the decoupling approximation is valid.
The fast detector has longer lifetime for the Kondo case. When using $t_d=0.5$~meV, $\Gamma=\gamma=1$~meV, $V_q=0.01$~meV, and $\Delta=0.01$~meV,
we obtain $\tau\sim 0.65~\mu$s for the Kondo detector.

\section{Discussions and coclusions}
We have shown that by measuring the shifts of the Kondo resonance peak 
and the Fano-Kondo dip in the conductance, we can estimate the optimal point and the tunneling strength
$\Omega$ between two states of a charge qubit. In general, it is believed that charged two-level systems are susceptible to phonons.
In Ref.~\onlinecite{tana0}, the result of the spin-boson model 
showed that the degradation of the coherence by phonons is smaller than expected.
Ref.~\onlinecite{Chen1,Chen2} also argued that the effect of phonons is not so large. 
In addition, because we use the coherent extrema, the effect of phonons is expected to be smaller 
than other energy scales.

We have studied the Kondo and the Fano-Kondo effects in QD system from viewpoint of the detectors 
of a capacitively coupled charge qubit. We have used the slave-boson mean field theory
and the decoupling approximation to describe the quantum interference of the system.
In particular, we have investigated the modulation of the conductance peak and dip 
by the charge qubit.
We found that, by measuring the shifts of the positions of the conductance peak and dip as a function of the applied gate voltage 
on the charge qubit (qubit bias),
we can estimate the optimal point. In addition, we showed that,  
by analyzing the derivatives of the shifts of the peak and dip as the function of the qubit bias,
we can infer the tunneling strength between the states $|0\rangle$ and $|1\rangle$ of the charge qubit. 
These characteristics are the results of the resonant behavior of the Kondo and the Fano-Kondo effects, 
and a new aspect of the application of these important quantum interference effects.

\acknowledgements
TT thanks A. Nishiyama, J. Koga and S. Fujita for useful discussions.
FN and XH are supported in part by NSA/LPS through ARO and DARPA QuEST through AFOSR.
FN was also partially supported by NSF grant
No. 0726909, JSPS-RFBR contract No. 09-02-92114,
Grant-in-Aid for Scientific Research (S), MEXT Kakenhi
on Quantum Cybernetics, and the JSPS via its FIRST program.
YXL is supported in part by the NNSFC under Grant Nos. 10975080 and 61025022.
\appendix

\section{Qubit-detector interaction}
\begin{figure}
\includegraphics[width=8cm]{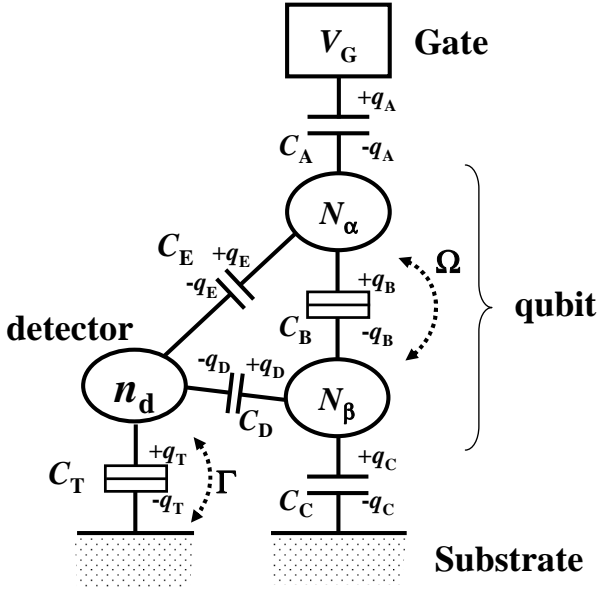}
\vspace*{8cm}
\caption{Charge qubit (right side) made of two coupled 
QDs with tunneling strength $\Omega$ and gate electrode $V_{\rm G}$. 
A detecting QD is in the left side and consists of a one-energy level 
state with tunneling coupling $\Gamma$ to reservoir (substrate). }
\label{AppFig1}
\end{figure}

Here we derive the formula of the capacitive interaction Eq.~(\ref{int}) between  a charge qubit  and a detector QD,
applying the capacitance network model to the system shown in Fig.~\ref{AppFig1}.
When charges stored in each capacitance are expressed as in Fig.~\ref{AppFig1}, 
the charging energy of this system is expressed by~\cite{tana1}
\begin{eqnarray}
\, U&=&\frac{q_A^2}{2C_A}+\frac{q_B^2}{2C_B}+\frac{q_C^2}{2C_C}+\frac{q_D^2}{2C_D}+\frac{q_E^2}{2C_E}+\frac{q_T^2}{2C_T}
\nonumber \\
&-& q_AV_G+q_T V_{\rm sub} +q_C V_{\rm sub} \ .
\end{eqnarray}
The numbers of electrons in the two QDs of the qubit and the detector QD site are described by the operators
$\hat{N}_\alpha$, $\hat{N}_\beta$ and  $\hat{n}_d$, respectively, such as
\begin{eqnarray}
\hat{N}_\alpha &\equiv& (-q_A+q_B+q_E)/e, \\
\hat{N}_\beta &\equiv& (-q_B+q_C+q_D)/e,  \\
\hat{n}_d &\equiv& (-q_E+q_D+q_T)/e. 
\end{eqnarray}
The charge distribution is determined by minimizing this charging energy.
When we define $\sigma_z \equiv \hat{N}_\alpha-\hat{N}_\beta$ under the 
condition $\hat{N}_\alpha+\hat{N}_\beta =1$, as in Ref.~\onlinecite{tana0} , we have 
\begin{eqnarray}
\hat{U} &=& \frac{C_t}{4D_z} \Bigl\{ 
 (\Theta_\beta-\Theta_\alpha) \sigma_z .
+2(C_\alpha C_\beta -C_{\rm B}^2)(\hat{n}_d-n_{d0})^2
\nonumber \\
& & +2 (C_p +C_m \sigma_z ) (\hat{n}_d-n_{d0}) \Bigr\}  + {\rm const.}\; ,
\label{U_trap}
\end{eqnarray}
where 
$C_\alpha\equiv C_A+C_B+C_E$, $C_\beta\equiv C_B+C_C+C_D$,
$C_t\equiv  C_D+C_E+C_T$, and $n_{d0}\equiv C_{\rm w}V_{\rm sub}$ 
with 
\begin{eqnarray}
\Theta_\alpha &\equiv& C_\alpha C_t-C_E^2, \ \Theta_\beta=C_\beta C_t-C_D^2, \\
D_z &\equiv & \Theta_\alpha \Theta_\beta -(C_DC_E +C_BC_t)^2 , \\
C_p&\equiv &-(C_\alpha+C_\beta)C_D -2C_\beta C_E \\
C_m&\equiv &C_\alpha C_D -C_\beta C_E +C_B (C_E-C_D) .
\end{eqnarray}

Thus, we can obtain the coupling $V_q$ between the charge qubit and the detector as 
\begin{equation}
V_q=\frac{C_tC_m}{2D_z} . 
\label{coupling0}
\end{equation}
When the detector is distant from the qubit, we can approximate $C_E=0$ and
$C_C\cong C_A$, and then we obtain
\begin{equation}
V_q\cong \frac{eC_D}{C_T(2C_B+C_A)} \propto \frac{1}{d_D}.
\label{coupling1}
\end{equation} 
When we can simply approximate the capacitance $C_D\approx \varepsilon S_D/d_D$ 
($\varepsilon$ is the dielectric constant, $S_D$ is the effective area of the QD, and 
$d_D$ is the distance between the qubit and the detector QD), 
we can see that the coupling constant is proportional to the inverse of 
the distance between the qubit and the detector QD, similar to pure Coulomb interaction.

\vspace*{1.5cm}



\end{document}